\documentstyle[aps,eqsecnum]{revtex}
\begin{document}
\title{Splitting Functions in the Light Front Hamiltonian Formalism}
\author{Rajen Kundu\thanks{e-mail : rajen@tnp.saha.ernet.in}\\
Saha Institute of Nuclear Physics, \\
1/AF Bidhannagar, Calcutta-700 064, India}
\date{17 May 1995}
\maketitle
\begin{abstract}
In this paper, we have calculated the lowest order splitting functions
occurring
in the Altarelli-Parisi equation in the light front hamiltonian formalism. Two
component perturbative LFQCD in light cone gauge is our guideline and we
have used LFTD-approximation for dressed quark and gluon states.
\end{abstract}
\section{Introduction}
It is believed that QCD is the potent theory for the correct description of
the strong interaction, but explaining the real situation like the hadrons
is still far from being complete. In the low energy non-perturbative sector,
it is plagued by the confinement and the vacuum structure problems among
others, which must be addressed properly to have the true picture of the
QCD bound state. Recently, there has been an increasing amount of interest
to explore the non-perturbative domain using the LFQCD in the light cone
gauge \cite{1}. Asymptotic freedom and the factorization
theorems which separates the soft and hard part of a QCD process, enables one
to use the perturbative treatment in the high energy sector. The soft part
measures the low energy (nonperturbative) properties of quarks and gluons
in the parent hadron and is connected to the parton distribution functions.
The hard part is relevant for the scale evolution of the hadronic structure
functions and we shall be mainly concerned with that in this paper.

It is a well known fact that the light front current commutators have the
bilocal structure in it and the fourier transform of such bilocal
matrix element gives the structure function of the deep inelastic scattering
\cite{2}. It is just one step further to see that parton pictures
emerge directly from it. Scaling of the structure functions is very
evident in this calculation. The systematic violation of the scaling comes
only when the QCD corrections are taken into account, as can be seen, for
example, from the Alteralli-Parisi equation \cite{3}.

Parton model is best realised in the infinite momentum
frame. Any kind of calculation involving parton model in the
equal-time framework assumes some infinite momentum limit (which is sometimes
conceptually difficult) and imposes some physical constraint on the gauge
field to get the meaningful result. We prefer to attack such problems using
light front framework and light cone gauge. Thus, with all the arsenal of
two-component perturbative LFQCD in the light cone gauge \cite{4}, we study the
problem from the first principle without any further assumption. In this paper
we calculate the lowest order splitting functions ($P_{qq}$,$P_{Gq}$,$P_{qG}$
and $P_{GG}$) \cite{3}
occurring in the AP-equation following this path.
This illustrates the efficiency and both conceptual and calculational
simplicity of perturbative LFQCD to produce some physical results.

In this calculation we have used the light front Tamm-Dancoff approximation
(LFTD) \cite{5}, to describe a dressed particle. This formalism
directly gives the probabilistic interpretation
of the splitting functions. Similar calculations of the splitting function
have been done in the equal time formalism taking infinite momentum limit in
the axial gauge  \cite{6}.

In this paper, we have briefly reviewed the necessary results in the first
part and we use those results in our calculation in the latter part of the
paper.
For detailed discussion one should consult the references mentioned within the
text.

\section{A Brief Review}
\subsection {LFQCD hamiltonian}
In our calculation we use two component formalism of LFQCD. Here we write
the LFQCD hamiltonian as a free term plus the interaction term:
\begin{equation}
H=\int dx^-d^2x_\perp ({\cal H}_0+{\cal H}_{int})
\end{equation}
where
\begin{eqnarray}
{\cal H}_0&=&{1\over 2}(\partial^iA^j_a)(\partial^iA^j_a)+\xi^\dagger({
-\partial^2_\perp+m^2\over i\partial^+})\xi\nonumber \\
{\cal H}_{int}&=&{\cal H}_{qqg}+{\cal H}_{ggg}+{\cal H}_{qqqq}+{\cal H}_{gggg}
\end{eqnarray}

Here ${\cal H}_{int}$ is divided into four parts depending on the nature of the
interaction which can be read off from the suffix, for example,
${\cal H}_{qqg}$ implies the interaction of two quarks with a gluon and so on.
First two terms of the interaction are of the order of $g$, the coupling
constant and the order of the other two is of $g^2$. Since, in our calculation
we shall only be concerned with the lowest order in the coupling constant,
we present here the first two terms explicitly.
\begin{eqnarray}
{\cal H}_{qqg}&=&g\xi^\dagger\left\{-2\left({1\over \partial^+}\right)(\partial
_{\perp}.A_{\perp})+(\sigma .A_{\perp})\left({1\over \partial^+}\right) (\sigma
{}.
\partial_{\perp}+m)+\left({1\over \partial^+}\right) (\sigma .
\partial_{\perp}-m)(\sigma.A_{\perp})\right\} \xi \\
{\cal H}_{ggg}&=&gf^{abc}\left\{(\partial^iA^j_a)A^i_bA^j_c+(\partial^iA^i_a)
\left({1\over\partial^+}\right)(A^j_b\partial^+A^j_c)\right\}
\end{eqnarray}

Here $A^i_a(x)$'s and $\xi(x)$ stands for the dynamical component of the gauge
fields and spinor fields. They are the solutions of the corresponding equation
of motion and given by the following expressions,
\begin{eqnarray}
A^i(x)&=&\sum_{\lambda}\int{dq^+d^2q^\perp\over 2(2\pi)^3[q+]}\left[\epsilon
^i_\lambda a(q,\lambda)e^{-iqx}+h.c\right] \\
\xi(x)&=&\sum_{\lambda}\chi_{\lambda}\int{dp^+d^2p^\perp\over 2(2\pi)^3}
\left[b(p,\lambda)e^{-ipx}+d^\dagger(p,-\lambda)e^{ipx}\right]
\end{eqnarray}
with $q^-={q_{\perp}^2\over q^+}$ and $p^-={p_{\perp}^2+m^2\over [p^+]}$.\\

Here $\lambda$ is defined to be the helicity,
\begin{eqnarray}
\lambda=\left\{ \begin{array}{ll}1 & \mbox{for gluons and}\\-1 & \mbox{ }
\end{array}\right.
\lambda=\left\{\begin{array}{ll}{1\over 2} & \mbox
{for quarks.}\\-{1\over 2} & \mbox{}\end{array}\right.
\end{eqnarray}
The gluon polarization vectors are $\epsilon^i_1={1\over\sqrt2}(1,i)$ and
$\epsilon^i_{-1}={1\over\sqrt2}(1,-i)$ and quark spinors are $\chi_{{1\over
2}}=\left(\begin{array}{ll}1 & \mbox{}\\0 & \mbox{}\end{array}\right)$,
$\chi_{-{1\over 2}}=\left(\begin{array}{ll}0 & \mbox{}\\1 & \mbox{}
\end{array}\right)$.
Also the creation and annihilation operators follow the basic commutation
relations
\begin{eqnarray}
[a(q,\lambda),a^\dagger(q^{\prime},\lambda^{\prime})]&=&2(2\pi)^3q^+\delta_
{\lambda\lambda^{\prime}}\delta^3(q-q^{\prime})\nonumber\\
\{b(p,\lambda),b^{\dagger}(p^{\prime},\lambda^{\prime})\}&=&\{d(p,\lambda),
d^{\dagger}(p^{\prime},\lambda^{\prime})\}=2(2\pi)^3\delta_{\lambda,\lambda
^{\prime}}\delta^3(p-p^{\prime})
\end{eqnarray}

In our calculations we also require the multiple principal value prescription,
given as
\begin{eqnarray}
\left({1\over \partial^+}\right)^nf(x^-)&=&\left({1\over
4}\right)^n\int^{\infty}_
{-\infty}dx^-_1dx^-_2....dx^-_n\epsilon(x^--x^-_1)....\epsilon(x^-_{n-1}-x^-_n)
f(x^-_n)\nonumber \\
&\longrightarrow&\left[{1\over 2}\left({1\over k^++i\epsilon}+{1\over k^+-i
\epsilon}\right)\right]^nf(k^+)\nonumber \\
&=&{1\over [k^+]^n}f(k^+)
\end{eqnarray}

\subsection{Two component perturbative LFQCD}
Here we present the rules that we have used to calculate the necessary matrix
elements.\\
a) Draw all topologically distinct $x^+$-ordered diagrams.\\
b) For each vertex, include a factor of $16\pi^3\delta(p_f-p_i)$ and a simple
matrix element given below. Each gluon line connected to the vertex
contributes a factor ${1\over \sqrt{k^+}}$ from the normalization of the
single gluon state.

Necessary matrix elements are as follows.
\begin{eqnarray}
H_{qqg}&\equiv&-gT^a_{\beta\alpha}\chi^{\dagger}_{\lambda_1}\left\{{2k^i\over
 [k^+]}-{(\sigma.p_2^{\perp}-im)\over [p_2^+]}\sigma^i-\sigma^i{(\sigma.p_1^
{\perp}+im)\over
[p_1^+]}\right\}\chi_{\lambda_2}\epsilon^{i*}_{\lambda}\nonumber\\
&\equiv&-gT^a_{\beta\alpha}\chi^{\dagger}_{\lambda_1}\Gamma^i(p_1,p_2,k)\chi_
{\lambda_2}\epsilon^{i*}_{\lambda} \\
H_{ggg}&\equiv&-igf^{abc}\epsilon^i_{\lambda_1}\epsilon^{j*}_{\lambda_2}
\epsilon^{l*}_{\lambda_3}\left\{\left[(k_2-k_3)^i-{k^i_1\over [k_1^+]}(k^+_2-
k^+_3)\right]\delta_{jl}\right.\nonumber \\
& &\left.+\left[(k_3+k_1)^j-{k^j_2\over [k_2^+]}(k^+_3+k^+_1)
\delta_{il}\right]+\left[-(k^1+k^2)^l+{k_3^l\over [k_3^+]}(k^+_1+k^+_2)
\right]\delta_{ij}\right\}\nonumber \\
&\equiv&-igf^{abc}\Gamma^{ijl}(k_1,k_2,k_3)\epsilon^i_{\lambda_1}\epsilon^{j*}_
{\lambda_2}\epsilon^{l*}_{\lambda_3}
\end{eqnarray}
Using energy-momentum conserving $\delta$-function and expressing them in
terms of relative momentum defined as $x={k^+\over [p^+]}$, $\kappa_i=k_i-
{k^+\over [p^+]}p_i$, we can write
\begin{eqnarray}
\Gamma^i(p,p-k,k)&=&2{k^i\over [k^+]}-{\sigma^j(p^j-k^j)-im\over [p^+-k^+]}
\sigma^i-\sigma^i{\sigma^jp^j+im\over [p^+]}\nonumber\\
&=&{1\over [p^+][1-x]}\left\{{2\over
[x]}\kappa^i-\sigma^i(\sigma.\kappa_{\perp})+i\sigma^imx\right\}\\
\Gamma^{ijl}(p,k,p-k)&=&\left[(p-2k)^i-{p^i\over [p^+]}(p^+-2k^+)\right]\delta
_{jl}\nonumber\\
& &+\left[(k-2p)^j-{k^j\over [k^+]}(k^+-2p^+)\right]\delta_{il}\nonumber\\
& &+\left[(p+k)^l-{p^l-k^l\over
[p^+-k^+]}(p^++k^+)\right]\delta_{ij}\nonumber\\
&=& 2\left\{-\kappa^i\delta_{jl}+{1\over [x]}\kappa^j\delta_{il}+{1\over [1-x]}
\kappa^l\delta_{ij}\right\}
\end{eqnarray}

\subsection{LFTD approximation in (3+1) dimension}
Here we briefly review the light front Tamm-Dancoff (LFTD) approximation in
(3+1) dimensions which will be necessary for our purpose. LFTD approximation,
in its simplest form, is just the Tamm-Dancoff approximation applied to light
front field theory. In Tamm-Dancoff approximation we usually truncate the
fock space as much as necessary and plausible, and then expand the dressed
particle state in the truncated fock space. For example, we can write the
state corresponding to a quark as follows:
\begin{eqnarray}
|quark\rangle&\equiv&|\psi(p^+,p^{\perp})\rangle\nonumber\\
&\equiv&{c_1(p)\over \sqrt
{2(2\pi)^3}}b^{\dagger}(p,\lambda)|0\rangle\nonumber\\
& &+\sum_{\lambda_1\lambda_2}\int
{dq^+d^2q^{\perp}\over \sqrt{2(2\pi)^3}}\int{dk^+d^2k^{\perp}\over \sqrt
{2(2\pi)^3}k^+}\delta^3(p-q-k)b^{\dagger}(q,\lambda_1)a^{\dagger}(k,\lambda_2)
|0\rangle c_2(q,k)\nonumber\\&=&c_1\psi_1+c_2\psi_2\;({\rm say})
\end{eqnarray}
[Here $\lambda$'s stand for spin as well as the colour indices.]

Thus we have truncated the fock space to a bare quark plus a quark and a
gluon state, so that the maximum occupation number is two.
For higher approximation we can take a three particle state also. Here $c_1$
and $c_2$ are the probability amplitude of
finding a bare quark with all the momentum and of finding a non-interacting
pair of quark-gluon with given momentum sharing.

Similarly, we can write a gluon as,
\begin{eqnarray}
|gluon\rangle&\equiv&|\psi(p^+,p^{\perp})\rangle\nonumber\\
&\equiv&{c_1(p)\over \sqrt
{2(2\pi)^3}p^+}a^{\dagger}(p,\lambda)|0\rangle\nonumber\\
& &+\sum_{\lambda_1\lambda_2}\int
{dq^+d^2q^{\perp}\over \sqrt{2(2\pi)^3}}\int{dk^+d^2k^{\perp}\over \sqrt
{2(2\pi)^3}}\delta^3(p-q-k)b^{\dagger}(q,\lambda_1)d^{\dagger}(k,\lambda_2)
|0\rangle c_2(q,k)\nonumber\\
& &+{1\over 2}\sum_{\lambda_1\lambda_2}\int
{dq^+d^2q^{\perp}\over \sqrt{2(2\pi)^3q^+}}\int{dk^+d^2k^{\perp}\over \sqrt
{2(2\pi)^3k^+}}\delta^3(p-q-k)a^{\dagger}(q,\lambda_1)a^{\dagger}(k,\lambda_2)
|0\rangle c_2^{\prime}(q,k)) \nonumber \\
&\equiv& c_1|bare-gluon\rangle+c_2|quark-antiquark\rangle
+c_2^{\prime}|gluon-gluon\rangle.
\end{eqnarray}

We also have to use within this fock space some momentum cutoff and in
general we should have a step function such that the invariant mass be
less than the cutoff.

One last comment about the normalization of the state. It is to be noted
that the state is not yet normalized. Using the normalization condition
one can get rid of one of the $c$'s, for example, $c_1(p)$.

\section{Splitting functions}
\subsection{Calculation of $P_{qq}$ and $P_{Gq}$}
First consider the state of a dressed quark according to the Tamm-Dancoff
approximation.
\begin{eqnarray}
|\psi(p^+,p^{\perp})\rangle&\equiv&{c_1(p)\over \sqrt
{2(2\pi)^3}}b^{\dagger}(p,\lambda)|0\rangle+\sum_{\lambda_1\lambda_2}\int
{dq^+d^2q^{\perp}\over \sqrt{2(2\pi)^3}}\int{dk^+d^2k^{\perp}\over \sqrt
{2(2\pi)^3k^+}}\delta^3(p-q-k)\nonumber\\
& &\;\;\;\;\;\;\;\;\;\;\;\;\;b^{\dagger}(q,\lambda_1)a^{\dagger}(k,\lambda_2)
|0\rangle c_2(q,k)\nonumber\\
&=&c_1\psi_1+c_2\psi_2\;({\rm say})
\end{eqnarray}
Here $\lambda $'s stand for spin as well as the colour indices.

Now, we consider the dispersion relation $(P^-_0+P^-_v)={M^2+P^2_{\perp}\over
P^+}$, and operate this on the quark state.
\begin{equation}
(P^-_0+P^-_v)|\psi\rangle={M^2+P^2_{\perp}\over P^+}|\psi\rangle
\end{equation}

Now take the projection of this equation onto non-interacting one-fermion
and one-fermion-one-boson state. This leads to
\begin{eqnarray}
\langle p^{\prime}|(P^-_0+P^-_v)|\psi\rangle&=&\langle p^{\prime}|
{M^2+P^2_{\perp}\over P^+}|\psi\rangle\nonumber\\
\langle p^{\prime}k^{\prime}|(P^-_0+P^-_v)|\psi\rangle&=&\langle p^{\prime}
k^{\prime}|{M^2+P^2_{\perp}\over P^+}|\psi\rangle\label{b}
\end{eqnarray}
where,
\begin{eqnarray}
& &|p^{\prime}\rangle\equiv{1\over \sqrt{2(2\pi)^3}}b^{\dagger}(p^{\prime}
,\lambda_1^{\prime})|0\rangle\nonumber\\
& &|p^{\prime},k^{\prime}\rangle\equiv{1\over \sqrt{2(2\pi)^3}}{1\over
\sqrt{2(2\pi)^3k^{+\prime}}}b^{\dagger}(p^{\prime},\lambda_1^{\prime})
a^{\dagger}(k^{\prime},\lambda^{\prime})|0\rangle
\end{eqnarray}

Now, from equation (\ref{b}), we get, up to the lowest order in $g$,
\begin{eqnarray}
& &c_2\langle p^{\prime}k^{\prime}|P^-_0|\psi_2\rangle+c_1\langle
p^{\prime}k^{\prime}|P^-_v|\psi_1\rangle=c_2\langle p^{\prime}k^
{\prime}|{M^2+P^2_{\perp}\over P^+}|\psi_2\rangle.\nonumber\\
{\rm or,}\;\;& &c_2\left[\langle p^{\prime}k^{\prime}|{M^2+P^2_{\perp}\over
P^+}
|\psi_2\rangle -\langle p^{\prime}k^{\prime}|P^-_0|\psi_2\rangle\right]
=c_1\langle p^{\prime}k^{\prime}|P^-_v|\psi_1\rangle.\nonumber\\
{\rm or,}\;\;& &c_2\left[{M^2+(p^{\prime}_{\perp}+k^{\prime}_{\perp})^2\over
(p^{\prime+}+k^{\prime+})}-{m^2_F+p^{\prime2}_{\perp}\over p^{\prime+}}-
{m^2_G+k^{\prime2}_{\perp}\over k^{\prime+}}\right]=c_1\langle p^{\prime}
k^{\prime}|P^-_v|\psi_1\rangle.\label{c}
\end{eqnarray}
Now,$\langle p^{\prime}k^{\prime}|P^-_v|\psi_1\rangle $ can easily be
calculated using LFQCD perturbation theory. We can use the rule already
mentioned
and take $M=m_F=m_G=0$ limit of the above expression. Also, take the total
transverse momentum to be zero.
\begin{equation}
\Rightarrow\;p^{\prime}_{\perp}+k^{\prime}_{\perp}=0\;{\rm or,}\;p^{\prime}
_{\perp} =-k^{\prime}_{\perp}=k_{\perp}\;(say)
\end{equation}
Thus, we parametrize the momentum as follows:
\begin{eqnarray}
{\rm total\; mom.}\;p&\equiv&(p,p,0_{\perp})\nonumber\\
{\rm quark\; mom.}\;p^{\prime}&\equiv&
\left((1-x)p+{k^2_{\perp}\over 2(1-x)p},(1-x)p,k_{\perp}\right)\nonumber\\
{\rm gluon\; mom.}\;k^{\prime}
&\equiv&\left(xp+{k^2_{\perp}\over 2xp},xp,-k_{\perp}\right).
\end{eqnarray}
With the above prescription, we can write equation (\ref{c}) as,
\begin{eqnarray}
c_2\left[-k^2_{\perp}\left({1\over x}+{1\over 1-x}\right)\right]&=&c_1\langle
 p^{\prime}k^{\prime}|P^-_v|\psi_1\rangle\nonumber\\
\Rightarrow c_2^2{k^4_{\perp}\over x^2(1-x)^2p^2}&=&c_1^2|\langle p^{\prime}
k^{\prime}|P^-_v|\psi_1\rangle|^2={c_1^2\over 2(2\pi)^3k^{\prime+}}
|\langle p^{\prime}k^{\prime}|P^-_v|p\rangle|^2.\label{d}
\end{eqnarray}
[Note that in the last step, we have written normalization factors and the
factor coming out of energy momentum conserving $\delta$-function separately,
so that the matrix element is now a number.]
Now, to get the correct $c_2^2$, we sum over all possible intermediate
states as we do for the case of an inclusive process and we get,
\begin{eqnarray}
c_2^2{k^4_{\perp}\over x^2(1-x)^2p^2}&=&{c_1^2\over 2(2\pi)^3k^{\prime+}}
g^2\sum_{a\beta}\left(T^a_{\beta\alpha}T^a_{\beta\alpha}\right)\chi^\dagger_
{\lambda_1}\Gamma^i(p,p^{\prime},k^{\prime})\sum_{\lambda_2}\chi_{\lambda_2}
\chi^{\dagger}_{\lambda_2}\Gamma^{\dagger j}\chi_{\lambda_1}\sum_{\lambda}
\epsilon^{i*}_{\lambda}\epsilon^j_{\lambda}\nonumber\\
&=&{c_1^2\over 2(2\pi)^3k^{\prime+}}
g^2\Gamma^i(p,p^{\prime},k^{\prime})\Gamma^{\dagger i}(p,p^{\prime},k^{\prime}
)\left({N^2-1\over 2N}\right)
\end{eqnarray}
where,
\begin{eqnarray}
\Gamma^i&=&{1\over [p^+][1-x]}\left\{{2k^i\over [x]}-\sigma^i(\sigma^j
k^j_\perp)\right\}\\
\Gamma^{\dagger i}&=&{1\over [p^+][1-x]}\left\{{2k^i\over [x]}-(\sigma^jk^j_
\perp)\sigma^i\right\}.
\end{eqnarray}
Here we have used,
\begin{eqnarray}
\sum_{a\beta}\left(T^a_{\beta\alpha}T^a_{\beta\alpha}\right) &=& {N^2-1\over
2N}\\
\sum_{\lambda_2}\chi_{\lambda_2}\chi^{\dagger}_{\lambda_2} &=& 1\\
\sum_{\lambda}\epsilon^{i*}_{\lambda}\epsilon^j_{\lambda} &=& \delta_{ij}.
\end{eqnarray}
Now,
\begin{eqnarray}
\Gamma^i\Gamma^{i\dagger}&=&{1\over p^2(1-x)^2}\left\{{4k^2_{\perp}\over x^2}
-{4(\sigma^jk^j_{\perp})^2\over x}+2(\sigma^jk^j_{\perp})^2\right\}\nonumber\\
&=&{2k^2_{\perp}\over x^2(1-x)^2p^2}\left[1+(1-x)^2\right]\\
\Rightarrow c_2^2(x,k_{\perp}) & = & {c_1^2\over 16\pi^3p}g^2\left({N^2-1\over
 2N}\right)
{1+(1-x)^2\over x}{2\over k^2_{\perp}},
\end{eqnarray}
as $k^{\prime+}=xp$.
Integrating over $k^{\perp}$, we get,
\begin{eqnarray}
c_2^2(x) &=& {c_1^2\over 16\pi^3p}g^2\left({N^2-1\over 2N}\right)
{1+(1-x)^2\over x}2\pi\ln\left({\Lambda\over \mu}\right)^2\nonumber\\
 &=& {c_1^2\over p}{\alpha\over 2\pi}P_{Gq}\ln\left({\Lambda\over \mu}
\right)^2\label{i}
\end{eqnarray}
where,
\begin{equation}
P_{Gq}(x)=\left({N^2-1\over 2N}\right){1+(1-x)^2\over x}\label{p}
\end{equation}
and $\alpha={g^2\over 4\pi}$. $\Lambda$ and $\mu$ are some ultraviolet and
infrared cutoff respectively.

$P_{qq}$ can be found using the relation
\begin{equation}
P_{qq}(x)=P_{Gq}(1-x)=\left({N^2-1\over 2N}\right){1+x^2\over 1-x}\label{t}.
\end{equation}
\subsection{Calculation of $P_{qG}$ and $P_{GG}$}
Now, if we want to calculate $P_{qG}$ or $P_{GG}$, we have to start with
a gluon state which is given by,
\begin{eqnarray}
|gluon\rangle&\equiv&|\psi(k)\rangle\nonumber\\
&\equiv&{c_1(k)\over \sqrt
{2(2\pi)^3}p^+}a^{\dagger}(k,\lambda)|0\rangle\nonumber\\
& &+\sum_{\lambda_1\lambda_2}\int
{dq^+d^2q^{\perp}\over \sqrt{2(2\pi)^3}}\int{dk^+d^2k^{\perp}\over \sqrt
{2(2\pi)^3}}\delta^3(p-q-k)b^{\dagger}(q,\lambda_1)d^{\dagger}(k,\lambda_2)
|0\rangle c_2(q,k)\nonumber\\
& &+{1\over 2}\sum_{\lambda_1\lambda_2}\int
{dk^+_1d^2k_1^{\perp}\over \sqrt{2(2\pi)^3k_1^+}}\int{dk_2^+d^2k_2^{\perp}\over
\sqrt{2(2\pi)^3k_2^+}}\delta^3(k-k_1-k_2)a^{\dagger}(k_1,\lambda_1)a^
{\dagger}(k_2,\lambda_2)|0\rangle c_2^{\prime}(k_1,k_2)\label{e}
\end{eqnarray}
Then we proceed along the same way as before. Consider the eigenvalue equation
for the gluon state and take the projection onto a quark-antiquark
state and two-gluon state separately.
\begin{eqnarray}
& &|quark-antiquark\rangle\equiv|p^{\prime}q^{\prime}\rangle\equiv{1\over
\sqrt{2(2\pi)^3}}b^{\dagger}(p^{\prime},\lambda_1^{\prime})d^{\dagger}
(q^{\prime},\lambda_2^{\prime})|0\rangle\\
& &|two-gluon\rangle\equiv|k_1^{\prime},k_2^{\prime}\rangle\equiv{1\over
\sqrt{2(2\pi)^3k_1^{+\prime}}}{1\over \sqrt{2(2\pi)^3k_2^{+\prime}}}a^{\dagger}
{\dagger}(k_1^{\prime},\lambda_1^{\prime})
a^{\dagger}(k_2^{\prime},\lambda_2^{\prime})|0\rangle
\end{eqnarray}
In the first case, 1st and 2nd term from the gluon state will contribute
and we get as in equation (\ref{d}),
\begin{equation}
c_2^2{k^4_{\perp}\over x^2(1-x)^2k^2}=c_1^2|\langle p^{\prime}
q^{\prime}|P^-_v|\psi_1\rangle|^2={c_1^2\over 2(2\pi)^3k}
|\langle p^{\prime}q^{\prime}|P^-_v|k\rangle|^2.
\end{equation}
Here, we have used the following momentum parametrization.
\begin{eqnarray}
{\rm total\; gluon\; mom.}\;k&\equiv&(k,k,0_{\perp})\nonumber \\
{\rm antiquark\; mom.}\;q^{\prime}&\equiv&\left((1-x)k+
{k^2_{\perp}\over 2(1-x)k},(1-x)k,-k_{\perp}\right)\nonumber \\
{\rm gluon\; mom.}\;p^{\prime}&\equiv&\left(xk+{k^2_{\perp}\over
2xk},xk,k_{\perp}\right).
\end{eqnarray}
Now, with this parametrization $\langle p^{\prime}q^{\prime}|P^-_v|k\rangle$
is exactly identical to $\langle p^{\prime}q^{\prime}|P^-_v|p\rangle$ we
already have calculated, with the difference that $\Gamma^i$ now becomes
\begin{eqnarray}
\Gamma^i&=&{1\over [k^+][1-x]}\left\{2k^i-{\sigma^i(\sigma^j
k^j_{\perp})\over [x]}\right\}\\
\Gamma^{i\dagger}&=&{1\over [k^+][1-x]}\left\{2k^i-{(\sigma^jk^j_
{\perp})\over [x]}\sigma^i\right\}
\end{eqnarray}
 Now, summing over all possible intermediate states, we get,
\begin{eqnarray}
\sum|\langle p^{\prime}q^{\prime}|P^-_v|k\rangle|^2&=&g^2\sum_{\beta\alpha}
\left(T^a_{\beta\alpha}T^a_{\beta\alpha}\right)\sum_{\lambda_1\lambda_2}
\left(\chi^{\dagger}_{\lambda_1}\Gamma^i\chi_{\lambda_2}\chi^{\dagger}_{\lambda_2}
\Gamma^{j\dagger}\chi_{\lambda_1}\right)\epsilon^{i*}_\lambda\epsilon^j_{\lambda}
\nonumber\\&=&g^2\left({1\over
2}\right)2\Gamma^i\Gamma^{j\dagger}\epsilon^{i*}_
\lambda\epsilon^j_{\lambda}
\end{eqnarray}
[The factor 2 comes from the summation over $\lambda_1$. ]

At this point we assume that the result is independent of initial gluon
polarizations and hence we add the result for different polarization and
divide by two to get the final result.
\begin{equation}
\rightarrow\sum|\langle p^{\prime}q^{\prime}|P^-_v|k\rangle|^2={1\over 2}
g^2\Gamma^i\Gamma^{i\dagger}
\end{equation}
[ As $\sum_{\lambda}\epsilon^{i*}_{\lambda}\epsilon^j_{\lambda}=\delta_
{ij}$. ]\\
Now,
\begin{equation}
\Gamma^i\Gamma^{i\dagger}={1\over k^2(1-x)^2}\left\{4k_{\perp}^2-{4(\sigma.
k_{\perp})^2\over x}+{2(\sigma.k_{\perp})^2\over x^2}\right\}=
{2k^2_{\perp}\over k^2x^2(1-x)^2}[x^2+(1-x)^2].
\end{equation}
\begin{equation}
\Rightarrow\;c_2^2(x,k_{\perp})={c^2_1\over 16\pi^3k}{1\over
2}g^2[x^2+(1-x)^2]{2\over
k^2_{\perp}}
\end{equation}
Integration over $k_{\perp}$ gives,
\begin{eqnarray}
c^2_2={c^2_1\over k}P_{qG}{\alpha\over 2\pi}\ln\left({\Lambda\over \mu}
\right)^2\label{j}
\end{eqnarray}
where,
\begin{equation}
P_{qG}={1\over 2}[x^2+(1-x)^2]\label{q}.
\end{equation}

In the second case, when we take the projection onto two-gluon state, 1st
and 3rd term in equation (\ref{e}) will contribute, and we get as in equation
(\ref{d}),
\begin{eqnarray}
c_2^{2\prime}{k^4_{\perp}\over x^2(1-x)^2k^2}&=&c_1^2|\langle k_1^{\prime}
k_2^{\prime}|P^-_v|\psi_1\rangle|^2\nonumber\\
&=&{c_1^2\over 2(2\pi)^3}{1\over k^3(1-x)x}
|\langle k_1^{\prime}k_2^{\prime}|P^-_v|k\rangle|^2.
\end{eqnarray}
[Again, we have extracted the factors arising from the normalization of
states and the energy-momentum conserving $\delta$-function, and write
them separately so that the remaining factor is just a number.]

Now, summing over all the intermediate gluon states and averaging over the
initial gluon states, we get the following result.
\begin{eqnarray}
\overline{\sum}|\langle q,k|P^-_v|p\rangle|^2&=&g^2
\left(\sum_{bc}f^{abc}f_{abc}
\right)\sum_{\begin{array}{lll}
i & j & l \\ i^{\prime} & j^{\prime} & l^{\prime}
\end{array}}\Gamma^{ijl}\Gamma^{\dagger
i^{\prime}j^{\prime}l^{\prime}}\nonumber\\
& &\;\;\;\;\;\;\;\;\;\;\;
{1\over
2}\sum_{\lambda_1}\epsilon^{i*}_{\lambda_1}\epsilon^{i\prime}_{\lambda_1}
\sum_{\lambda_2}\epsilon^{j*}_{\lambda_2}\epsilon^{j\prime}_{\lambda_2}
\sum_{\lambda_3}\epsilon^{l*}_{\lambda_3}\epsilon^{l\prime}_{\lambda_3}
\end{eqnarray}
Now using the relations
\begin{eqnarray}
\sum_{bc}f^{abc}f_{abc}&=&N\\
\sum_{\lambda_2}\epsilon^{i*}_{\lambda_2}\epsilon^{i\prime}_{\lambda_2}&=&
\delta_{ii^{\prime}}.
\end{eqnarray}
\begin{equation}
\Rightarrow\overline{\sum}|\langle q,k|P^-_v|p\rangle|^2=g^2 N{1\over
2}\sum_{ijl}
\Gamma^{ijl}\Gamma^{ijl\dagger}
\end{equation}
Now,
\begin{equation}
\Gamma^{ijl}\Gamma^{ijl\dagger}=8k_{\perp}^2\left[1+{1\over x^2}+{1\over
(1-x)^2}
\right].
\end{equation}
\begin{equation}
\rightarrow \overline{\sum}|\langle q,k|P^-_v|p\rangle|^2=g^2 N
\left[1+{1\over x^2}+{1\over (1-x)^2}\right]4k_{\perp}^2
\end{equation}
Thus, using this result, we get,
\begin{equation}
c^{2\prime}_2(x,k_{\perp})={c_1^2\over k}\left[x(1-x)+{x\over 1-x}+{1-x\over
x}\right]
(2N)\left({g^2\over 16\pi^3}\right)\left({2\over k^2_{\perp}}\right).
\end{equation}
Integration over $k_{\perp}$ gives,
\begin{equation}
c^{2\prime}_2={c^2_1\over k}P_{GG}{\alpha\over 2\pi}\ln\left({\Lambda\over
\mu}\right)^2\label{k}
\end{equation}
where,
\begin{equation}
P_{GG}=2N\left[x(1-x)+{x\over (1-x)}+{1-x\over x}\right]\label{r}.
\end{equation}
\section{conclusion}
 Using the normalization condition we can write $c^2_1=1$, in the equations
 (\ref{i}), (\ref{j}), and (\ref{k}), since we are working only with the lowest
order
 contributions
 in the coupling constant. Thus, we can see directly that the splitting
 functions are proportional to the corresponding probabilities as is the case
 in AP-equation.
 Equation (\ref{p}), (\ref{t}), (\ref{q}) and (\ref{r}) show the expressions
for
 the splitting functions which are
 exactly identical to that derived in AP-paper.

Of course, we have not considered the contributions coming from the infrared
singularity ($x\rightarrow 1$ limit) in the expression of splitting function.
One usually puts them on the basis of some physical arguments or by direct
loop calculations.

 In principle one should be able starting from the first principle, to derive
 the AP-equation in the hamiltonian framework. The very success of our
 calculation in obtaining the splitting functions suggests that it may be
plausible to try
 and extrapolate it to obtain AP-equation, which
 is, of course, a non-trivial job.

\acknowledgements

This paper owes heavily for its idea and the constant help in various parts
of the calculation to Prof. Avaroth Harindranath, SINP, Calcutta. I also
acknowledge the valuable help that I have got from the members of the TNP
division of the institute.


\begin{references}
\bibitem {1} K. G. Wilson {\it et al}, Phys. Rev., {\bf D49}, 6720 (1994).
\bibitem {2} J. M. Cornwall and R. Jackiw, Phys. Rev., {\bf D4}, 367 (1971);
   D. A. Dicus, R. Jackiw
   and L. Teplitz, Phys. Rev., {\bf D4}, 1733 (1971); T. M. Yan, Phys. Rev.,
   {\bf D4}, 1760 (1973).
\bibitem {3} G. Alteralli and P. Parisi, Nuc. Phys., {\bf B126}, 298 (1977).
\bibitem {4} For the development of the theory, see W. M. Zhang and A.
Harindranath,
   Phys. Rev., {\bf D48}, 4868; {\bf D4}, 4881 and {\bf D4}, 4903 (1994).
\bibitem {5} R. Perry and A. Harindranath, Phys. Rev., {\bf D43}, 4051 (1992).
Also see
   S. J. Brodsky and G. P. Lepage, {\it Exclusive Processes in QCD}, edited by
   A. H. Mueller (World Scientific, Singapore, 1989).
\bibitem {6} W. Greiner and A. Sch\"afer, {\it Quantum Chromodynamics},
 Springer (1994).
\end{references}
\end{document}